\documentclass[11pt]{article}
\usepackage{Blois,epsfig}

\def\Journal#1#2#3#4{{#1} {\bf #2}, #3 (#4)}

\def\NPB{{\em Nucl. Phys.} B}
\def\PLB{{\em Phys. Lett.}  B}
\def\PRL{\em Phys. Rev. Lett.}

\def\be{\begin{equation}}
\def\ee{\end{equation}}
\def\bea{\begin{eqnarray}}
\def\eea{\end{eqnarray}}

\newenvironment{e}{\begin{equation}}{\end{equation}}

\begin{document}
\vspace*{2cm}
\begin{center}
\Large{\textbf{XIth International Conference on\\ Elastic and Diffractive Scattering\\ Ch\^{a}teau de Blois, France, May 15 - 20, 2005}}
\end{center}

\vspace*{2cm}
\title {ELASTIC AND INCLUSIVE INELASTIC DIFFRACTION\\ OF HIGH ENERGY HADRONS}

\author{ Andrzej R. MA{\L}ECKI }

\address{Instytut Fizyki AP, 
{ul. Podchor{\c{a}}{\.z}ych 2,\\ PL 30-084 Krak{\.o}w}, Poland}

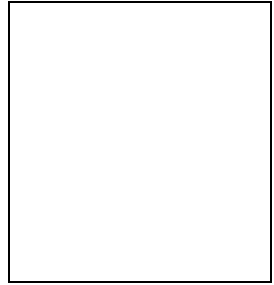
\maketitle\abstracts{ \hspace{.7cm}  In the classical approach
the inelastic diffraction originates from the diversity of
elastic scattering amplitudes in the initial and final states. We
consider a multi-channel correction, accounting for intermediate
transitions inside an equivalence class of diffractive states. It
can be factorized in the form $N \Delta t$, to be taken in the
Bjorken-like 
 limit: $N\rightarrow \infty,\Delta
t\rightarrow 0$ such that $N\Delta t$ is finite.
     Our formalism provides an excellent description of elastic and inclusive inelastic
scattering of high energy (53-546 GeV) protons and antiprotons.}


\section {Diffractive limit of hadronic collisions}
\hspace{.7cm}
 There are striking analogies between
light scattering and high-energy collisions of hadrons which
consist in a substantial presence of elastic scattering due to a
feed-back from inelastic channels. The optical resemblances of
high-energy hadron diffraction should not, however, be
overemphasized. In fact, the 'diffractive structure' of the
hadronic elastic differential cross-section is  often obscured
since multiple dips and reinforcements may not be present. The
hadron 'inelastic diffraction', involving transitions with no
exchange of intrinsic quantum numbers, has no classical analogy
at all. It appears as a peculiar quantum phenomenon related to
the existence of internal degrees of freedom.

 \hspace{.7cm}The most convenient basis for calculating the diffractive
  amplitude is provided by the experimentally revealed division of inelastic
  channels into the diffractive and non-diffractive transitions.
  The decomposition of the space of physical states
  (with respect to the initial state) into subspaces of
  diffractive $[D]$ and non-diffractive states $[\sim D]$ implies
  the existence of unitary operators $U$ and $U^{\dag}$
  which are block-diagonal in the Hilbert space:
    $\langle k \mid Uj\rangle =\langle k \mid U^{\dag}j \rangle  =
  0$   for any $|j \rangle \in [D]$ and $|k \rangle \in [\sim D]$.
   Expanding the initial $|i\rangle$ and final
$|f\rangle$ states in the basis of $|Uj\rangle$ states one
obtains the amplitude of diffractive transitions in terms of the
matrix elements of the normal operator $\Lambda \equiv 1 - U$ :
 \begin{e}
  T_{f\, i}\: =
t_{i}\delta_{f\, i}\: - N_{fi}(T_0)\Lambda_{fi}t_{i}\: -
\Lambda_{if}^{\star}t_{f}N_{if}^{\star}(T_0 ^{\dag})\: + \:
\sum_{|j\rangle \in [D]} N_{fj}(T_0)\Lambda_{fj}
t_{j}\Lambda_{ij}^{\star}
\end{e}
where  $\Lambda_{kj}\equiv \langle k\mid \Lambda\mid j\rangle $,
$t_{kj}\equiv \langle Uk\mid T\mid Uj\rangle$, $t_{j} \equiv
t_{jj}$ being the diagonal matrix elements of $T_0 \equiv
{U}^{\dag} T U$, while the undimensional quantities
\begin{e}
N_{kj}(T_0)\equiv\frac{1}{\Lambda_{kj}t_{j}}\sum_{|l\rangle\in
[D]}\Lambda_{kl} t_{lj}
\end{e}
 allow to reduce the summations
\cite{ARM}. If the subspace [D] contains a very large number of
diffractive states then $ N_{kj}\equiv N\rightarrow\infty $ for
any pair of states $|k\rangle$ and $|j\rangle $. In fact, since $
\Lambda $ is a non-singular operator its matrix elements vary
smoothly under the change of diffractive states. This leads to an
enormous simplification of Eq.(1) in the limit
$N\rightarrow\infty $:
\begin{e}
\label{TfiN}
 T_{f\, i}\: =\: t_{i}\delta_{fi}\:
-N(\Lambda_{fi}t_{i}+\Lambda_{if}^{\star}t_{f}-\sum_{|j\rangle
\in [D]}\Lambda_{fj}t_{j}\Lambda_{ij}^{\star}).
\end{e}
\section {Elastic diffraction}
\hspace{.7cm}
   In general, the effect of non-diagonal transitions inside
 the diffractive subspace [D] gets factorized.
  E.g., in the case of elastic scattering one has:
\newline
\begin{e}
 \label{TiiN} T_{ii}\:= \:t_{i}\: + \: N \sum_{|j\rangle
\in [D]}{|\Lambda_{ij}|}^2 (t_{j} - t_{i})\:=\: t_{i}\: +\: g_{i}
N (t_{av}^{(i)} - t_{i})
\end{e}
where
  $ g_{i} = \sum_{|j\rangle}{|\Lambda_{ij}|}^2\: = 2
Re(\Lambda_{ii}) $ and
$t_{av}^{(i)}= \frac{1}{g_{i}} \sum_{|j\rangle}{|\Lambda_{ij}|}^2
t_{j}$
 is the average value of the diagonal matrix elements $t_{j}$.
 The expressions of the form $ N \Delta t $ where $\Delta t $
represents diversity of $ t_{j} $ over the subspace of
diffractive states [D] are to be considered in the {\em
diffractive limit} \cite{ARM}: $ N\rightarrow\infty $, %
$ \Delta t \rightarrow 0 $ 
such that $ N \Delta t $  is finite.
 The  inelastic diffraction
contribution is thus built as an {\em infinite sum} of the {\em
infinitesimal} contributions from all intermediate states
belonging to [D].

\hspace{.7cm} Our numerical analysis of elastic and inelastic
scattering was done in the framework of a model where the
diffractive states are built of a two-hadron core (representing
the ground state) and some quanta describing diffractive
excitations \cite{ARM}:
$|j\rangle=|i\rangle+\;|n;\vec{b}_{1}\;.\;.\;.
\vec{b}_{n}\rangle$. The configurations of these quasi-particles
(called \emph{diffractons}) are specified by a number $\emph{n}$
of constituents and their impact parameters
$\vec{b}_{1},\;.\;.\;.,\vec{b}_{n}$. Thus %
$\frac{1}{g_{i}}\sum_{|j\rangle\in[D]}|\Lambda_{ij}|^{2}\ldots=
\sum_{n=1}^\infty P_{n}\int d^{2}b_{1}\ldots
d^{2}b_{n}\prod_{k=1}^{n} |\Psi(b_{k})|^{2}\ldots $ where
$|\Psi(b_{k})|^{2}$ is the density of a spatial distribution of
diffractons (with respect to the core) in the impact plane and
$P_{n}$ are probabilities of their number, approximated by
Poisson distributions. The diagonal matrix elements of $T_{0}$
(in $b$-space) are specified in terms of the real profile
functions. We have
 $N (t_j -t_i ) =i
(1-\Gamma_0)\lim_{N\rightarrow\infty ,\gamma\rightarrow 0}N
\sum_{k=1}^n \gamma (\vec{b}-{\vec{b}_k} )$ with
$t_{i}=i\Gamma_{0}$ representing the hadronic core and $\gamma$'s
corresponding to diffractons. The diffracton model thus
explicitly accounts for the \emph{geometrical} diffraction on an
absorbing hadronic bulk and the \emph{dynamical} diffraction
corresponding to intermediate transitions between diffractive
states.

 \hspace{.7cm}   Our analysis, though based on the
well-founded theoretical framework, has a semi-phenomenological
character. The shapes of the profiles $\Gamma_0$, $\gamma$ and of
the density ${|\psi (b)|}^2 $ are assumed, for simplicity, as
Gaussians. Their parameters, as well as the coupling constant
$g_i$ and the mean number of diffractons $\langle n\rangle$, were
determined from fitting to experimental data (Fig. 1). Analogous
excellent fits were performed to the experimental data on
proton-antiproton elastic scattering at the c.m. energies $\sqrt
s$= 53 GeV \cite{Break}, $\sqrt s$= 546-630 GeV \cite{Bozzo} and
1800 GeV \cite{Amos}.

 For high energies we always had:
$\sigma_0>>\sigma_{n}$ and $R_0>R_{n}$. This is reasonable since
the non-diffractive effects dominate a long-range part of
scattering and are characterised by large values of the effective
coupling strength. The diffractive scattering, on the other hand,
is governed by short distance dynamics and small values of the
coupling strength.
 \pagestyle{plain} \setcounter{figure}{0}
\begin{center}
\begin{figure}
\vspace{40 mm}
 \includegraphics{el53.ugs}
 \caption [.] {At small momentum
transfers the non-diffractive contribution (dashed curve) is
dominant. The diffractive (dotted) term has a single zero which,
filled up by the real part of the scattering amplitude, appears
as a shallow minimum. Above the dip the non-diffractive
contribution is negligible and the diffractive term dominates the
elastic cross-section. The solid curve results from the sum of
the two contributions in Eq. \ref{TfiN}. It was fitted to 44
experimental points from the data on proton-proton elastic
scattering at the
c.m. energy $\sqrt s $=52.8 GeV \cite{Schub}. The values 
found are $\sigma_0$ = 39.4 mb, $R_0$ = 0.70 fm,
    $\sigma_{n}$ = 5.52 mb,  $R_{n} $ = 0.41 fm,
     $\rho\equiv Re T_{ii}(0)$/$Im T_{ii}(0)$ = 0.066.}
\end{figure}
\end{center}

\section {Inclusive inelastic diffraction}

\hspace{.7cm} Making use of completness of diffractive states in
the equivalence subspace one may obtain from (\ref{TfiN}) the
inclusive cross-section of inelastic diffraction:

\begin{e}
\label{Tfi2N'} \sum_{|f\rangle\neq |i\rangle}{|T_{fi}|}^2 \: =\:
N^2 \sum_{|f\rangle \in
[D]}{|\Lambda_{if}|}^2{|t_{av}^{(i)}-t_{f}|}^2 \: +\: ({1-g_{i}})
{g_{i}}N^2\,{\mid t_{av}^{(i)}-t_{i}\mid}^2
\end{e}
 The  inelastic diffraction is thus
built of the two contributions: an incoherent one which is
proportional to a dispersion of the $T_0$-diagonal matrix
elements and the coherent contribution which is proportional to
the square of diffractive term in the elastic scattering
amplitude.

   The name of incoherent contribution is justified by its proportionality,
in a leading order, to the mean value $\langle n\rangle$. It
appears in the form of the double Fourier-Bessel transform:
\begin{e}
\label{incoh}
\frac{d\sigma_{incoh}(q)}{d^2q}=\frac{1}{(2\pi)^2}\int
d^2b_1d^2b_2 e^{i{\vec q\cdot (\vec b_1-\vec
b_2)}}[1-\Gamma_0(b_1)][1-\Gamma_0(b_2)]I({\vec b_1,\vec b_2})
\end{e}
where the function
$\label{Ib1b2} I({\vec b_1,\vec b_2})=N^2 g_i\langle n\rangle
U({\vec b_1,\vec b_2})$
depends on the correlation function of diffractons
$U({\vec b_1,\vec b_2})\equiv\int d^2s|\psi(s)|^2\gamma({\vec b_1-\vec s})\gamma({\vec b_2-\vec s})$.

\setcounter{figure}{1}
\begin{figure}
\vspace{40 mm} \includegraphics{inel53.ugs}
\caption[.]{The coherent and incoherent contributions to
inclusive inelastic diffraction. At small momentum transfers the
coherent contribution (dotted curve) is dominant. At the momentum
transfer which corresponds to the position of the dip in elastic
differential cross-section the coherent contribution becomes
negligible and the incoherent term (dashed curve) dominates the
inelastic diffraction at large momentum transfers. The solid
curve which is the sum of the two contributions was fitted to 30
points of p - p inelastic diffraction at
 $\sqrt s$= 53 GeV \cite{AlbArm}.}
\end{figure}

 For the
description of inclusive inelastic diffraction only 3 parameters:
$g_i,\langle n\rangle$ and the diffracton radius $R_{\epsilon}$
are required since the remaining parameters are to be determined
from elastic scattering.
 The angular distributions of
inelastic diffraction measured at the ISR and SPS colliders
\cite{AlbArm,Bernard} are, in a wide range of energy,
consistently characterised by two different slopes at small and
large momentum transfers. The experimental results could
therefore be well reproduced simply with a sum of two Gaussians
described by 4 parameters: two slopes and two other parameters
which fix the forward magnitude of each Gaussian. However, in our
phenomenology we need only 3 parameters since the slope at small
momentum transfers is already determined by the diffractive term
in elastic scattering. The strength of this term in inelastic
diffraction is set-up by the coupling constant $g_{i}$. In
elastic scattering this constant was hidden in the definition of
the cross-section $\sigma_n$ and in inelastic diffraction it
appears as a new parameter at disposal.

    It should be stressed that the coherent contribution 
to the inclusive cross-section is a novelty of our approach.
    We  claim that the shape of inelastic diffraction at small
momentum transfers is determined by elastic scattering in the
transition region between the forward peak and the diffraction
minimum.  As Figs 2 and 3 show, this is successfully verified in
experiment \cite{AlbArm,Bernard}, being an important 
evidence in favor of our formalism.

 \setcounter{figure}{2}
\begin{figure}
\vspace{60 mm}
 \includegraphics{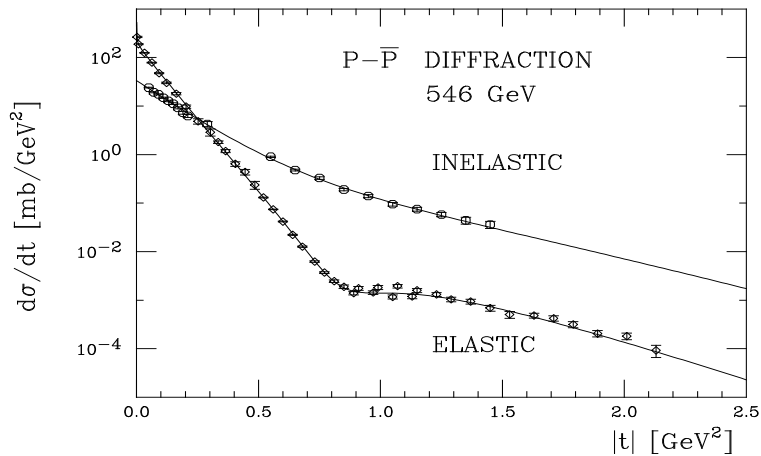}
 \caption [.]{Elastic and
inclusive inelastic diffraction at 546 GeV \cite{Bozzo,Bernard}.}
\end{figure}

\end{document}